# Measuring Presence in Augmented Reality Environments: Design and a First Test of a Questionnaire

**Holger Regenbrecht**

DaimlerChrysler Research and Technology

Ulm, Germany

regenbre@igroup.org

**Thomas Schubert**

Friedrich-Schiller-University

Jena, Germany

schubert@igroup.org

## Introduction

Augmented Reality (AR) enriches a user's real environment by adding spatially aligned virtual objects (3D models, 2D textures, textual annotations, etc) by means of special display technologies. These are either worn on the body or placed in the working environment. From a technical point of view, AR faces three major challenges: (1) to generate a high quality rendering, (2) to precisely register (in position and orientation) the virtual objects (VOs) with the real environment, and (3) to do so in interactive real-time (Regenbrecht, Wagner, & Baratoff, in press). The goal is to create the impression that the VOs are part of the real environment. Therefore, and similar to definitions of virtual reality (Steuer, 1992), it makes sense to define AR from a psychological point of view: Augmented Reality conveys the impression that VOs are present in the real environment. In order to evaluate how well this goal is reached, a psychological measurement of this type of presence is necessary. In the following, we will describe technological features of AR systems that make a special questionnaire version necessary, describe our approach to the questionnaire development, and the data collection strategy.

Finally we will present first results of the application of the questionnaire in a recent study with 385 participants.

## 1. Technology and Type of Presence

AR systems can be categorized regarding the display and interaction technology used as well as the reference relationships between real world, virtual (augmented) world, and the user's body or body parts (esp. hands). In addition to that one should distinguish between single and multi user applications. Widely used are systems with an HMD as a display device, either as an optical-see-thru system, where the user is looking through the glasses of the HMD into the real environment, or video-see-thru systems, where the real world is mediated by a video stream captured by a head-mounted mini camera. The whole system can be used in a stationary setup or as a wearable technique (Azuma, 1997). A second category of AR systems uses (stereo) projection techniques (e.g. CAVE-like environments, stereo projection screens) in addition to real world objects (e.g. Bimber et. al., 2001). A third category uses hand-held devices with a Through-the-lens metaphor to augment VOs onto the real environment seen through the display (Regenbrecht & Specht, 2000, Mogilev et. al., 2002).

Although from a technical perspective these systems are very different, they share a common feature, namely that it is the real environment that the VOs are placed in. Therefore, the common approach to measure presence in virtual environments, questions on the experience whether one has the sense of being there in the VE, does not work. AR elicits a different sense of presence: "It is here" presence (Lombard & Ditton, 1997). We know of no previous attempts to develop a scale for this experience.

## 2. Questionnaire Development

We have developed a first version of a questionnaire to measure the experienced presence of VOs in the real environment. Our approach was based on a previous theoretical model of presence in virtual environments that is easily generalizable to presence of VOs in augmented reality (Schubert, Friedmann, & Regenbrecht, 2001). We argue that presence of VOs develops when we mentally represent bodily actions on the VOs as *potential bodily actions*. In order to develop these representations, it is necessary to devote attention to them and actively construe them on the basis of their visual representation. Related to the experience that the VOs are present in the real space, but by no means identical to it, is the experience that they achieve a certain sense of realness (Banos et al., 2000). In our previous research, we have found that presence and realness judgments differ, and that they are predicted by different variables (Schubert, Friedmann, & Regenbrecht, 1999).

On the basis of this reasoning, we developed items that assessed (a) the experienced presence of VOs in the real space, (b) experienced co-location of VOs and body in the same space, (c) experienced realness of the VOs, (d) synaesthetic experiences and behavioral confusion, as well as (e) experienced control over the interaction and (f) experienced effort for mental construal. We also added items on previous experience with similar technologies and enjoyment of the interaction. After a first round of data collection with the technologies mentioned below (N=16), items with very low variance were excluded. Although preliminary, these data already suggested that the division between presence and realness also holds for AR.

## 3.     Data collection

The final questionnaire, with 26 questions related to facets (a) through (f), was recently applied to users of four different applications, out of each of the realms described above: 1) HMD-based single user application "MagicDesk" (Regenbrecht, Baratoff, & Wagner, 2001), 2) the interactive two user Through-the-lens demonstration "AR Pad" and HMD-based "MagicMeeting" (Regenbrecht & Wagner, 2002), and 3) the single/multi user projection system "IllusionHole" at Bauhaus University Weimar, and 4) the special projection system "Virtual Showcase" using the "Raptor" application. These applications feature the following interactions: *MagicDesk*: In front of an HMD-wearing user stands a turnable plate (CakePlatter) on a table with a virtual scaled car model on it. Using a pointing device with a knob on it the participants can texture car parts (like door or roof) by pointing a virtual ray towards the desired part and pushing the button. The texture to be used currently is "photographed" with a knob on the HMD before. *AR Pad*: Two user, each holding a LCD screen with a camera attached to it. Mounted to this device is a "SpaceOrb" device for 3DOF rotation and selection. The users sit vis-a-vis on a table. Between the users textured cubes are placed in virtual (augmented) space. Each side of a cube is textured different. The task is to puzzle in 3D. *MagicMeeting*: This is the collaborative version of MagicDesk, in our case a two user version. The users sit side by side on the same table looking onto one shared model with collaborative interactions. The interaction tasks are lighting the model, a real transparent acryl clipping plane (virtually clipping the model) and annotation cards to color the model. *IllusionHole*: Between one and four users look at a large horizontal computer monitor screen, wearing shutter glasses. The position of their heads are tracked. Each user sees a specific part of the screen, on which the VOs are presented correctly aligned for his or her perspective. The partition of the screen space is reached by a hole in a plane placed above the screen. The hole is a circle, with a diameter of approximately 20 cm. The VOs appear to flow in the space left by the hole, and extend up to 10 cm below and above the hole. Participants can interact with the objects using tracked pens. *Raptor*: One or more users are standing around a table with a half-silvered mirror cone on top of it. Within the cone a real (exhibition) object is placed, in this case the skull of a dinosaur. This object is lit with projectors in a very special technological way, so that some parts (like muscles) can be spatially augmented. The exhibition was supplemented by audio narration.

In each environment, users interact at least 5 minutes with the environment. Because of the very simple and natural interfaces no detailed instructions are needed. Participants can use the systems immediately. We currently collect data using the questionnaire and will present results from principal components analysis and regressions on the other measures at the conference. We will especially discuss (a) the relation between "it is here" presence and realness, (b) the relation to the experienced control over the interaction, and (c) the similarities and differences to presence in virtual environments.

## 4. Raptor Study

The first comprehensive application of the questionnaire developed took place at the conference / fair SIGGRAPH 2002. We describe this survey a little bit more in detail, although the questionnaire was applied in a very special setup: the Virtual Showcase Raptor environment.

### Participants

In total, 385 participants filled in the questionnaire. The experimenter conducting the study marked 19 of them as invalid due to insufficient language abilities or non-serious participation. Additional 13 participants were excluded due to missing values in the factor-analysed variables. Thus, 353 participants remained in the analysed sample. Of them, 80.5% were male. Mean age was 34.8.

### Method

The participants were wearing tracked 3D shutter glasses to experience the augmented 3D space. After a couple of minutes of exploring the Raptor model (skull with augmented muscles) with recorded audio explanations they filled in the questionnaire on the experiences and attitudes toward the system. Of importance to the present purposes, there were 7 items on presence-related experiences, 3 items on previous experiences with virtual reality, artificial reality, and computer games, an item on how comfortable the HMD was, an item on whether the recorded audio narration was helpful, and 5 items on the acceptance of the system (whether they would try a similar technology again, whether they deemed the technology as suitable for a museum, whether they would be willing to pay a higher entrance fee, how much they would pay in addition, whether they would prefer to go to such an exhibition). All items were answered on 7 point Likert type items.

### Factor Analysis

The sample was suitable for a principal component analysis, as indicated by a KMO score of .70. Three components had an Eigenvalue above 1. A first strong component explained 30.9% of the variance, the second (15.9%) and third (14.68%) component explained considerably less variance. We nevertheless extracted 3 components since a test with 2 components showed that these 2 did not form coherent scales, but that the items of the third components would again fall out. The components were obliquely rotated (Direct Oblimin).

Table 1 shows the structure matrix. Component 1 combines items related to how real the virtual objects seemed and how well they integrated with the real objects. Component 2 combines 2 items on the 3dness of the virtual objects. Component 3 taps experiences of the perceptual process itself - whether the difference between real and virtual drew attention, and whether the perception of the virtual object was needed effort. We call the first component *realness*, the second component *spatial presence*, and the third component *perceptual stress*.

|  | Component | | |
| --- | --- | --- | --- |
|  | 1 | 2 | 3 |
| P3 Was watching the virtual objects just as natural as watching the real world? | .746 | .292 |  |
| P2 Did you have the impression that the virtual objects belonged to the real object (dinosaur skull), or did they seem separate from it? | -.745 |  | .228 |
| P4 Did you have the impression that you could have touched and grasped the virtual objects? | .686 | .346 | -.126 |
| P5 Did the virtual objects appear to be (visualized) on a screen, or did you have the impression that they were located in space? | .187 | .828 |  |
| P6 Did you have the impression of seeing the virtual objects as merely flat images or as three-dimensional objects? | .271 | .801 | -.203 |
| P7 Did you pay attention at all to the difference between real and virtual objects? | -.220 |  | .785 |
| P8 Did you have to make an effort to recognize the virtual objects as being three-dimensional? |  | -.318 | .714 |

*Table 1: Component structure matrix*

We then explored the relation between these presence factors and the acceptance of the AR system. For components 1 and 2, mean scores were computed (Alphas = .56 and .60, respectively). Component 3 was dropped since the Alpha (.27) and the intercorrelation of the 2 items ($r=.16$) was insufficient.

Furthermore, we created an average score of the acceptance items (Alpha=.65). We then regressed the acceptance score on realness and spatial presence, previous experiences with AR, VR, and computer games, the comfort of the HMD, helpfulness of the audio narration, age, and gender. Stepwise regression was used. The final regression model explained a significant amount of variance, $F(4,303)=18.97$, $p<.001$ (lower N due to missing values). Significant positive predictors were realness, spatial presence, narration helpfulness, and HMD comfort, $t$s > 2.4, $p$s<.017. All experience scores, age, and gender dropped out.

Discussion

In the component analysis of items assessing presence of augmented reality, we found distinct factors for the experienced realness of the virtual objects, and the spatial presence of these objects. Whether the virtual objects are experienced as three-dimensional and in space is distinct from whether they seem to be real and well integrated with the real environment. The two components are correlated ($r=.299$), but distinct in the component analysis. Furthermore, it seems that perceptual stress comes out as a third factor, but the two items on it correlate not highly, making this a very preliminary finding. A regression analysis showed that both realness and spatial presence contribute to the acceptance of an AR system.

The distinction between realness and spatial presence reminds of previous results finding the same factor for immersive virtual environments (Schubert et al., 2001). There, the perception of the VE as surrounding the body (spatial presence) and the perception of the VE as real were also distinct from each other. The distinction becomes even more important in AR, where real objects have to be perceptually integrated with virtual objects.

## 5. Acknowledgments

We'd like to thank Michael Wagner, Mark Billinghurst, Oliver Bimber, Bernd Fröhlich, and Miguel Encarnacao for their support.

## 6. References

Azuma, R (1997). A Survey of Augmented Reality. Presence:Teleoperators and Virtual Environments, 6(4), 355-385.

Banos, R. M., Botella, C., Garcia-Palacios, A., Villa, H., Perpina, C., Alcaniz, M. (2000). Presence and reality judgment in virtual environments: A unitary construct? CyberPsychology and Behaviour, 3(3), 327-335.

Bimber, O., Fröhlich, B., Schmalstieg, D., and Encarnacao, L.M. (2001). The Virtual Showcase, IEEE Computer Graphics and Applications, 21(6), 48-55.

Lombard, M., & Ditton, T. At the heart of it all: The concept of presence. Journal of Computer-Mediated Communication, 3(2), http://www.ascusc.org/jcmc/vol3/issue2/lombard.html.

Mogilev, D., Kiyokawa, K., Billinghurst, M., and Pair, J. (2002). AR Pad: An Interface for Face-to-Face AR Collaboration. Extended Abstract in Proceedings of CHI 2002, April 20-25, 2002, Minneapolis, Minnesota, USA. ACM, 654-655.

Regenbrecht, H., Baratoff, G., & Wagner, M. (2001). A tangible AR desktop environment. Computers & Graphics, 25(5), 755-763.

Regenbrecht H.T. & Specht, R. (2000). A mobile Passive Augmented Reality Device - mPARD. Short Paper at International Symposium on Augmented Reality IEEE ISAR2000, Munich/Germany, October 5-6, 2000.

Regenbrecht, H. & Wagner, M.(2002). Interaction in a Collaborative Augmented Reality Environment. Extended Abstract in Proceedings of CHI 2002, April 20-25, 2002, Minneapolis, Minnesota, USA. ACM, 504-505.

Regenbrecht, H., Wagner, M., & Baratoff, G. (in press). MagicMeeting - a Collaborative Tangible Augmented Reality System. Virtual Reality - Systems, Development and Applications, 6(3).

Schubert, T., Friedmann, F., & Regenbrecht, H. (1999). Embodied Presence in Virtual Environments. In Ray Paton & Irene Neilson (Eds.), Visual Representations and Interpretations (pp. 269-278). London: Springer-Verlag.

Schubert, T., Friedmann, F., & Regenbrecht, H. (2001). The experience of presence: Factor analytic insights. Presence: Teleoperators and Virtual Environments, 10(3), 266-281.

Steuer, J. S. (1992). Defining virtual reality: Dimensions determining telepresence. Journal of Communication, 42(4), 73–93.